\documentclass[12pt]{article}
\usepackage{graphicx}
\usepackage{amsfonts}
\usepackage{amsmath}
\usepackage{enumerate}
\input epsf
\usepackage[pdftex,
            pdfauthor={A.A. Yurova, A.V. Yurov and V.A. Yurov},
            pdftitle={And now for something completely different: what the anthropic principle can tell us about the future of the dark energy universe},
            pdfsubject={},
            pdfkeywords={Anthropic principle, dark energy, cosmological constant},
            pdfproducer={Latex with hyperref},
            pdfcreator={pdflatex}]{hyperref}
\usepackage{color,soul}
\usepackage{xcolor}
\hypersetup{colorlinks=true, citecolor=blue, linkcolor=red}
\newcommand{\beq}{\begin{equation}}
\newcommand{\beqn}{\begin{equation*}}
\newcommand{\enq}{\end{equation}}
\newcommand{\enqn}{\end{equation*}}

\newcommand{\R}{{\mathbb R}}

\renewcommand{\L}{\Lambda}

\newcommand{\footremember}[2]{%
    \footnote{#2}
    \newcounter{#1}
    \setcounter{#1}{\value{footnote}}%
}
\newcommand{\footrecall}[1]{%
    \footnotemark[\value{#1}]%
}

\begin{document}
\font\bss=cmr12 scaled\magstep 0
\title{What the anthropic principle can tell us about the future of the dark energy universe}
\author{Alla A. Yurova \footremember{Kant}{Immanuel Kant Baltic Federal University, Institute of Physics, Mathematics and Informational Technology,
 Al.Nevsky St. 14, Kaliningrad, 236041, Russia} \footnote{Kaliningrad State Technical University, Department of Advanced Mathematics, Sovetsky Avenue 1, Kaliningrad, 236000, Russia} \footnote{AIUrova@kantiana.ru} \and Artyom V. Yurov \footrecall{Kant} \footnote{AIUrov@kantiana.ru}
 \and Valerian A. Yurov\footrecall{Kant} \footnote{vayt37@gmail.com}}%
\date {}
\maketitle
\begin{abstract}
An anthropic explanation for evident smallness of the value of the dark energy implies the existence of a time-dependent component of the scalar field, serving, together with a negative-valued cosmological constant, as one of two components to the overall density of the dark energy. The observers (i.e. us) might then only evolve in those regions of the universe where the sum of those two components (a positive and a negative ones) is sufficiently close to zero. However, according to Vilenkin and Garriga, the scalar field component has to slowly but surely diminish in time. In about a trillion years this process will put a cap to the now-observable accelerated expansion of the universe, leading to a subsequent phase of impending collapse. However, the vanishing scalar field might also produce some rather unexpected singularities for a finite non-zero scale factor. We analyse this possibility on a particular example of Sudden Future Singularities (SFS) and come to a startling conclusion: the time required for SFS to arise must be ``comparable'' to the lifetime of observable universe.
\end{abstract}
\thispagestyle{empty}
\medskip

\section{Introduction} \label{Sec:Intro}

In a seminal work \cite{Garriga_Vilenkin} by Garriga and Vilenkin a very novel ``anthropic'' approach has been presented purporting to resolve two enigmas surrounding the ``cosmological constant'': its smallness and its time coincidence. The key idea there was to view the density of the dark energy $\rho_{_D}$ as a random variable. To be more precise, $\rho_{_D}$ was to be written as a sum of two terms:
\begin{equation}
\rho_{_D}=\rho_{_{\Lambda}}+\rho_{_X},
\label{1}
\end{equation}
with $\rho_{_{\Lambda}} \in \R$ being the vacuum energy's density,
and $\rho_{_X}$ -- a variable density of a dynamical dark energy component $X$. This implies that $|\rho_{_\L}|$ doesn't have to be small as is usually accepted, for as long as $|\rho_{_{\Lambda}}| \sim |\rho_{_X}|$ and $\rho_{_{\Lambda}}\rho_{_X}<0$, then automatically $|\rho_{_D}| \ll 1$. This, coupled with the anthropic restrictions on the observable value of $\rho_{_D}$~\footnote{Interestingly, the very first paper which has suggested the anthropic solution of the cosmological constant due to the dark energy dates back to 1986 \cite{Linde-1}!}  (see \cite{Carter}--\cite{Weinberg}) provided a radically new view on the problem of the smallness of dark energy. In addition to that, the authors of \cite{Garriga_Vilenkin} have managed to find a very elegant way of explaining the time coincidence of the cosmological constant with the density of dark matter\footnote{According to the observational data the present densities of the dark matter (DM) and dark energy (DE) are of approximately the same magnitude. This ``coincidence'' is actually very puzzling because the evolution of DM is rather distinct from that of DE, and because there should be just a relatively narrow time frame within which their densities would coincide. And yet we somehow happen to be right in the midst of it!}, as well as to make a number of testable predictions (see also \cite{Linde-2}). One of them, being most exciting, but at the same time the hardest to verify, was the prediction that the expansion of the observable universe will, at some time in the future, stop, only to be replaced with a gravitational collapse (however, the earliest this can happen is after about a trillion years of accelerated expansion). Here is how this startling conclusion came to pass: first, it was assumed that the term $\rho_{_X}$ in (\ref{1}) was a contribution of some scalar field  ${\tilde \phi}$ with a very low mass \cite{2}, \cite{3}. On the next step, one recalls that the conscious observers (i.e. us) can only emerge within a very narrow spectrum of values of $\rho_{_D}$\footnote{Throughout this article we will be using the system of units where $8\pi G/3=c=1$.}:
\begin{equation}
-t^{-2}_{_{EI}}\le \rho_{_D}\le t^{-2}_{_{EG}},
\label{1.2}
\end{equation}
where $t_{_{EI}}$ is the (earliest) time required for intelligence to emerge and develop, and $t_{_{EG}}$ is the time it takes for the earliest galaxies to form (if $t_0$ is the present time then $t_{_{EG}}\sim \left(1+z_{_{EG}}\right)^{-3/2} t_0$ with redshift $z_{_{EG}}\sim 5$).

With the ``band of opportunity'' (\ref{1.2}) being extremely narrow, the potential $V({\tilde\phi})$ inside of it remains practically constant, and one can estimate that $V({\tilde\phi})\sim V(0)+V'(0){\tilde\phi}$. Therefore, if ${\tilde\phi}$ decreases in time, one is free to use the slow-roll approximation \cite{Liddle_Lyth}--\cite{YYY} to get an estimate on the time $t_*$ which measures how long it will take from the current time $t_0$ till the beginning of the collapse. In particular, if ${\tilde\phi}_0={\tilde\phi}(t=t_0)$ and $V(0)=0$, then $t_*\sim t_{_D} ({\tilde\phi}_0/m_p)^2$, where $t_{_D}\sim \rho{_{_D}}^{-1/2}$, and so we end up with $t_*\sim 10^{12}$ yr.

In this article we are going to describe another unusual prediction that can be made by studying the behaviour of the dynamical component of the dark energy. For that end, let us introduce a new field variable $\phi$, such that in terms of this variable (\ref{1}) can be formally rewritten as
\begin{equation}
\rho_{_D}=V(\phi).
\label{1.3}
\end{equation}
Then, let's assume that, in accordance with \cite{Garriga_Vilenkin}, that the field $\phi$ varies but very slowly, and that the way the effective potential in (\ref{1.3}) depends on $\phi$ can be approximately described by a power law:
\begin{equation}
V(\phi)\sim A^2\phi^n,
\label{1.4}
\end{equation}
where all the variables are real-valued -- including $n$, which in general is {\em not necessarily} an integer. Such potentials has been extensively studied in \cite{4-Bar} by Barrow and Graham, who drew one particularly interesting conclusion: the potentials (\ref{1.4}) might lead to singularities spontaneously emerging in finite time as $\phi\to 0$.

In particular, let $M$ be an integer such that
\begin{equation}
M < n < M+1,
\label{1.5}
\end{equation}
then as $\phi\to 0$ the scalar field develops a divergence in the $M+2$ derivative with respect to time: $d^{M+2}\phi/dt^{M+2}\to (-1)^{M+1}\times \infty$, while all the lower order derivatives remain finite. Such behaviour is characteristic of the class IV singularities by the Nojiri-Odintsov-Tsujikawa classification \cite{NOT}, and can be characterized as a ``weak'' singularity according to the terminology of \cite{Tipler}, \cite{Krolak-Polish}. For example, if $M=0$ then only the second order derivative $\ddot{\phi}$ will diverge, whereas both the density $\rho$ and the pressure $p$ remain finite.

The presence of such peculiar singularities in an otherwise very realistic field model with the with power-law potential leads to yet another interesting observation: the prediction by Garriga and Vilenkin might actually be a lot more verifiable then originally thought, since the emergence of those singularities happens much sooner than the aforementioned time scale $t_*\sim 10^{12}$ yr (see. Sec. \ref{Sec:Imminence}).

In order to get a grasp on the characteristic time that has to pass before the emergence of such a singularity, we will consider a simple integrable cosmological model. In Sec. \ref{Sec:Unforseen_Consequences} we will describe the method of superpotential that will be useful in determining the dynamics of singular solutions and will subsequently discover an even stronger singularity -- a Sudden Future Singularity (SFS) -- arising in finite time, provided the potential has the form:
\begin{equation}
V(\phi)=C_1\phi^n+C_2\phi^m,
\label{1.6}
\end{equation}
where $nm<0$. This observation, coupled with the fact that the potential (\ref{1.6}) is quite permissible in the cosmology, makes us to conclude that the Garriga-Vilenkin hypothesis can actually predicate an emergence of SFS. Following this line of reasoning, in Sec. \ref{Sec:Imminence} we actually build such a model under the approximation of slow-varying $\phi$. The detailed analysis of that model leads us to something quite unexpected: the conclusion that the time scale in which the SFS emerges has to be comparable to the lifetime of the observable universe! In other words, if SFS should happen, it will be very soon (that is, ``soon'' in cosmological terms, of course!).

\section{Some Singularities are Sudden} \label{Sec:Unforseen_Consequences}

We begin by pondering the potential (\ref{1.4}). In the flat Friedmann-Lema\^{i}tre-Robertson-Walker (FLRW) model the dynamics of the universe is given by the system:
\beq \label{FLRW}
\begin{split}
H^2 &= \frac{1}{2}{\dot\phi}^2+V(\phi)\\
\ddot\phi = &-3H\dot\phi - V'(\phi),
\end{split}
\enq
which in case of the power-law potential \eqref{1.4} turns into:
\begin{equation}
H^2=\frac{1}{2}{\dot\phi}^2+A^2 \phi^n,
\label{2.1}
\end{equation}
\begin{equation}
\ddot\phi=-3H\dot\phi-nA^2\phi^{n-1}.
\label{2.2}
\end{equation}

Let $\phi_0>0$, $H_0>0$ and $\dot\phi_0>0$. Then the r.h.s. of (\ref{2.2}) is strictly negative, i.e. $\ddot\phi<0$, and $\dot\phi$ shall decrease, eventually turning to zero. If that happens in finite time (which is easy to verify) then the scalar field $\phi$ itself will begin to decrease. And if $0< n <1$, then at some time $\phi\to 0$ and $\ddot\phi\to-\infty$. Despite this, $\dot\phi$ remains finite, and so do $\rho=H^2$ and $p=\dot\phi^2/2-V(\phi)$. In other words, we end up with a scale factor whose third (and higher) order derivative with respect to $t$ diverges at some moment of time (for more details cf. \cite{4-Bar}).

The potentials (\ref{1.4}) are generally non-integrable, which significantly complicates our explorations. Fortunately, we have in our possession a versatile tool called the {\em method of superpotential} \cite{superW}, which would allow us to overcome this difficulty by producing a suitable completely integrable model. Here is how it works.

First, we introduce a {\em superpotential} $W(\phi)$, defined as
\begin{equation}
W(\phi)=\frac{1}{2}{\dot\phi}^2+V(\phi).
\label{2.3}
\end{equation}
Using (\ref{2.3}), it is easy to see that the FLRW equations \eqref{FLRW} get reduced to a much simpler system:
\begin{equation}
\frac{d\phi}{dt}=-\frac{1}{3}\frac{W'(\phi)}{\sqrt{W(\phi)}},
\label{2.4}
 \end{equation}
\begin{equation}
V(\phi)=W(\phi)-\frac{1}{9}\frac{\left(W'(\phi)\right)^2}{W(\phi)}.
\label{2.5}
\end{equation}

The key to the next step would be to introduce the desired power-law relationship, except we do it {\em not} for potential $V(\phi)$, but for superpotential $W(\phi)$ instead. Then, letting $W(\phi)=A^2\phi^n$ and integrating (\ref{2.4}) we end up with:
\begin{equation}
\phi(t)=C\left(t_s-t\right)^{2/(4-n)}, \qquad 0 < n < 4
\label{2.6}
\end{equation}
where we have introduced an integration constant $t_s$, and
$$
C=\left(\frac{An(4-n)}{6}\right)^{\frac{2}{4-n}}.
$$
What kind of potential $V$ are we dealing with? This can be answered by looking at (\ref{2.5}):
\begin{equation}
V(\phi)=A^2\phi^n-\frac{A^2n^2}{9}\phi^{n-2},
\label{2.7}
\end{equation}
i.e. we have a potential that will be exactly like (\ref{1.6}), provided that $0 < n < 2$. If we set $2<n<4$, then the potential would still be (\ref{1.6}) but with a positive power $m>0$. It is now easy to see that

\begin{enumerate}[(i)]

\item If $0<n<2$, then $\dot\phi$ diverges as $t\to t_s$, i.e. $\dot\phi\to -\infty$;

\item If $2<n<3$, then $\ddot\phi\to\infty$ as $t\to t_s$, but $\dot\phi$ remains finite;

\item If $3<n<10/3$, then $\dddot\phi \to-\infty$  as $t\to t_s$, while both $\dot\phi$ and $\ddot\phi$ remain finite. Note, that since we are only interested in the non-integer powers of $\phi$ we are thereby restricting ourselves to strict inequalities.

\end{enumerate}

Naturally, when $n>2$ we end up with the very solutions that were discovered and described in \cite{4-Bar}. However, for the purpose of this article we will instead concentrate our efforts on the case (i), which is the only case that does not fit into the class of singularities studied in \cite{4-Bar}. The reason for not fitting in is obvious: the class (i) singularity is an SFS! We can verify it by directly calculating the density and pressure:
\begin{equation}
\rho=H^2=A^2C^n\eta^{\frac{2n}{4-n}},
\label{2.8}
\end{equation}
\begin{equation}
p=\frac{A}{3(4-n)}\eta^{-\frac{2(2-n)}{4-n}}\left(2nC^{n/2}-3AC^n(4-n)\eta^{\frac{4}{4-n}}\right),
\label{2.9}
\end{equation}
where $\eta=t_s-t$. So, as $t\to t_s$ we have $\rho\to 0$ and $p\to +\infty$ (again, we assume that $0<n<2$), which proves that the strong ($\rho+3p\ge 0$) and the weak ($\rho+p\ge 0$) energy conditions remain valid.

The revelation that SFS can be generated by the potentials like (\ref{1.6}) with $m<0$ (of which (\ref{2.7}) is but an example) actually should not come as too big of a surprise. It is easy to see that all polynomial potentials
\begin{equation}
V(\phi)=\sum_k c_k\phi^{n_k},
\label{3.1}
\end{equation}
with $n_k>0$ are incapable of producing SFS. Indeed, the requirement that $p\to \infty$, while $\rho\to \rho_s<\infty$ as $t\to t_s$, can be satisfied if and only if the kinetic term and the potential were to diverge at the same time albeit with opposite signs, whereas $\phi\to\phi_s<\infty$  (recall that by assumption $\phi_s=0$). But the potential (\ref{3.1}) with $n_k>0$ is regular at $\phi\to 0$. Accidentally, this explains why there were no SFS in \cite{4-Bar}.

Nevertheless, the SFS popping into existence by the vanishing scalar field poses a very interesting question: could this scenario be replicated in the Garriga-Vilenkin model by the slow evolution of the dynamical component of a scalar field? In other words, can it be that the very reason for the existence of the conscious observers -- almost complete compensation of the vacuum energy by the scalar field -- is something that might in close future render it inhospitable? That would certainly a case of a silver lining conjuring its own cloud! Let's look at this possibility in the next section.

\section{Is Collapse Imminent?}\label{Sec:Imminence}

Let's assume that we were right in our assumptions: that the anthropic model of Garriga-Vilenkin is correct and that there is a slowly vanishing scalar field which (eventually) produces SFS. Let us move a little further by assuming a more general class of SFS, where as $t\to t_s$ we have $p\to+\infty$, but $\rho\to\rho_s\ne 0$ \cite{Yurov}. As we have discussed in Sec. \ref{Sec:Intro}, the anthropic principle restricts the possible $\rho_{_D}$ to a very narrow band (\ref{1.2}), where $\rho_{_D}$ can be defined as (\ref{1.3}). This implies the following dynamics: $\rho_{_D}$ essentially behaves as a constant ($\rho_s$) up to the imminent arrival of SFS. The same can be said about the pressure $p_{_D}\sim -\rho_{_D}$, although at $t=t_s$ we have $p_{_D}\to +\infty$, unlike $\rho_{_D}$ who remains finite. Such a model can be easily constructed, if we assume that
\begin{equation}
\rho=\Lambda+\frac{C_{_{DM}}}{a^3},
\label{3.2}
\end{equation}
\begin{equation}
p=-\Lambda+\alpha^2_{_S}\delta\left(\frac{t_s-t}{T}\right),
\label{3.3}
\end{equation}
where $\delta$ is the Dirac delta-function. Before we move on, a remark would be in order, regarding this function. At a first glance it might seem counterintuitive to incorporate such an object into the pressure function, especially in view of SFS being a ``soft'' singularity. However, it is instructive to recall that the second derivative of a scale factor at the very moment of sudden singularity ceases to be an analytic function. The same can be said about the pressure, courtesy of the FLRW equation (recall that both $a$ and $\rho$ remain finite during SFS):
\beq
\label{Friedmann}
\ddot a = -\frac{1}{2}(\rho+3p) a.
\enq
Furthermore, it has been recently shown by Fern\'andez-Jambrina and Lazkoz \cite{Fernandez04}--\cite{Fernandez07} that the sudden singularities like those introduced in \cite{Barrow04} have one interesting property: any geodesic in a Friedmann universe do not end at the sudden singularity but safely pass through it. This implies that there might exist a unique description of the universe which unites the pre- and post-SFS cosmological dynamics as one unbroken evolution, passing through the sudden singularity. In such a framework a discontinuity in the pressure dynamics would appear as an infinitesimally thin but infinitely high jump in a pressure plotted as a function of time. This naturally leads us to a function akin to \eqref{3.3}. Finally, in order to fit the definition of SFS, we have to satisfy the following three conditions: at $t=t_s$ we should have \cite{bar-Bar}:
\beqn
\begin{split}
&0 < a(t_s) < \infty,\\
&0 <{\dot a}(t_s)<\infty,\\
&\lim_{t \to t_s}\ddot a = -\infty.
\end{split}
\enqn
As we shall see, {\em all} of these conditions are indeed satisfied by \eqref{3.3}, so we do have a bona fide sudden singularity.

Now let's return back to the model \eqref{3.2}--\eqref{3.3}. When $\alpha^2_{_S}=0$ our model morphs into a $\Lambda$CDM model, where $\rho_{_D}= \Lambda$, and $p_{_D}=p$, owing to the fact that there is no contribution to the pressure from the dark and baryon matters (note, that baryon are accounted for in the $C_{_{DM}}$ coefficient in (\ref{3.2}), being a small perturbation of the dark matter term). The system (\ref{3.2}), (\ref{3.3}) perfectly satisfies our requirements, so the corresponding equations can be used to qualitatively describe a very slow dynamics of $X$ component in the anthropic model of Garriga and Vilenkin. As such, we expect that the estimates we are about to make using \eqref{3.2}-\eqref{3.3} will not change in a significant way even for more difficult and more realistic models, provided those realistic models still satisfy the Garriga-Vilenkin postulates.

Let's now move on towards actually building those estimates. The solutions of FLRW equations will have a form:
\begin{equation}
a(t<t_s)=a_{-}(t)=a_0\left(\frac{\Omega_{_{DM}}(t_0)}{\Omega_{_\Lambda}(t_0)}\sinh^2\frac{t}{T}\right)^{1/3} = B\sinh^{2/3}\left(\frac{t}{T}\right),
\label{3.4}
\end{equation}
\begin{equation}
a(t>t_s)=a_{+}(t)=a_0\left(\frac{\Omega_{_{DM}}(t_0)}{\Omega_{_\Lambda}(t_0)}\sinh^2\frac{2t_s-t}{T}\right)^{1/3}= B\sinh^{2/3}\left(\frac{2t_s-t}{T}\right),
\label{3.5}
\end{equation}
where
$$
\Omega_{_\Lambda}(t_0)=\frac{\Lambda}{H^2(t_0)},\quad \Omega_{_{DM}}(t_0)=\frac{C_{_{DM}}}{a^3(t_0)H^2(t_0)},\quad T=\frac{2}{2\sqrt{\Lambda}}, \quad B= a_0\left(\frac{\Omega_{_{DM}}(t_0)}{\Omega_{_\Lambda}(t_0)}\right)^{1/3}.
$$
As we have discussed, we are interested in the dynamics of the universe where the scale factor and its first order derivative are both continuous functions everywhere. This implies that at $t=t_s$ we have to impose a matching condition for (\ref{3.4}), (\ref{3.5}):
\begin{equation}
a_{+}(t_s)=a_{-}(t_s)=B\sinh^{2/3}\tau_s,
\label{3.6}
\end{equation}
\begin{equation}
{\dot a_{+}}(t_s)=-{\dot a_{-}}(t_s)=-\frac{2B \cosh\tau_s}{3T(\sinh\tau_s)^{1/3}},
\label{3.7}
\end{equation}
where $\tau_s=t_s/T$. Here we again note that such ``stitching'' together of two seemingly different solutions (pre- and post-SFS) is not just a mathematical sleigh of hand, but a direct application of the results of Fern\'andez-Jambrina and Lazkoz \cite{Fernandez04}--\cite{Fernandez07} regarding a possible existence of unique description of a universe enduring the sudden singularity and surviving it. The matching condition \eqref{3.6}--\eqref{3.7} is then just a mathematical manifestation of this hypothesis.

Now, if we look at our model at $t=t_s$, we shall see that the density remains constant, the pressure blows up (a standard SFS behaviour), and the Hubble parameter changes sign:
\begin{equation}
H_{+}(t_s)=-H_{-}(t_s)=-\frac{2}{3T}\coth\frac{t_s}{T},
\label{3.8}
\end{equation}
so starting out from $t>t_s$ the universe enters a phase of collapse, instigated by the SFS at $t=t_s$.

We still have two undefined parameters, though: a positively defined ``SFS coupling constant'' $\alpha^2_{_S}$ and $t_s$ -- the moment of the emergence of SFS. We can  define them in a following way. First, we substitute (\ref{3.2}) and (\ref{3.3}) into the Friedmann equation \eqref{Friedmann}, then we integrate it inside of the interval $t\in (t_s-\epsilon,t_s+\epsilon)$ and take a limit $\epsilon\to 0$. Our reward would be the following important relationship:
\begin{equation}
\Lambda \coth\frac{t_s}{T}=\alpha^2_{_S}.
\label{3.10}
\end{equation}
If we choose $\Lambda\sim 0.7\times 10^{-29}$ $g/cm^{-3}$, then $T\sim 10$ Gyr. Since we are making a sensible assumption that SFS will happen in the future, then $t_s>T$ and we have an estimate:
\begin{equation}
1<\frac{\alpha^2_{_S}}{\Lambda}<\coth 1\sim 1.31.
\label{3.11}
\end{equation}

As the exact value of this fraction is arbitrary, we can use the Bayes approach (a.k.a. the mediocrity principle) and say that $\alpha^2_{_S}/\Lambda$ should lie somewhere around $1.155$, i.e. in the middle of interval (\ref{3.11}). If this is so, then the most probable estimate we can make would be
e\begin{equation}
t_s\sim T\times \coth^{-1} 1.155\sim 13.16\,\,\,\, {\rm Gyr}.
\label{3.12}
\end{equation}

Of course, this result has to be taken with a grain of salt, being just an approximation. Still, it looks very unlikely that the better estimates were to differ from $T$ by orders of magnitude. In fact, for those more accurate estimates to predict $t_s\gg T$, it would either require a presence of some hitherto unknown ``fine tuning'' of cosmological parameters, or indicate the work of some undiscovered physical laws behind the choice of $\alpha^2_{_S}$ --  the ``SFS coupling constant''. Both if these possibilities seem very unlikely (at least at this juncture) which makes us to believe that the result we have gathered is sufficiently general for those anthropic Garriga-Vilenkin models with a slowly vanishing scalar field that contain SFS.

\section{Conclusion}

Since the our final result is admittedly rather unusual, it would be a good idea to once again briefly go over the reasoning that had led us here.

\begin{enumerate}
 \item We began with the anthropic solution to the cosmological constant problem, developed in \cite{Garriga_Vilenkin}. Its core assumption was that the dark energy consists of two components: the vacuum energy and a slowly varying dynamical component. Then the density of the vacuum energy might actually satisfy the predictions imposed by the string theory: being large and negative-valued. As long as the density of a second, dynamical, component is sufficiently large and positively-defined, the anthropic principle will do the rest, i.e. select those universes where the sum of both components is significantly small to ensure the universe in question is habitable, thus producing the dark energy density that is compatible with the observational data.

\item On the other hand, the model of \cite{Garriga_Vilenkin} predicts in no uncertain terms that the dynamical component of the dark energy has to be slowly decreasing, thus predicting in a distant future a change in the dynamics of the universe. About $10^{12}$ years of accelerated expansion the universe is supposed to stop growing and begin to contract. If the variable component is produced by a scalar filed (another alternative would be a four-form field strength, which can vary through nucleation of membranes \cite{2}, \cite{BT}), then the dark energy density will play the role of an effective self-action potential of this very slowly abating scalar field.

\item It is sensible to assume that this potential should have a form  $V\sim \phi^n$. On the other hand, there are no reasons to a priori set constant $n$ to be an integer -- or even a positive number. But if $n$ is {\em not} an integer, then, according to \cite{4-Bar} the vanishing scalar field might give rise to some previously unforseen new singularities. In particular, since we cannot rule out the effective potentials (\ref{1.6}), the abating field might lead
    to SFS, a Sudden Future Singularity (see Sec. \ref{Sec:Unforseen_Consequences}). In what has followed, we have concentrated specifically on this type of singularity, with it being the most extreme and therefore more interesting (the ``weaker'' type of singularities will be considered in subsequent papers).

\item In order to study the emergence of SFS  in the Garriga-Vilenkin approach we have constructed a very simple intuitive model (see Sec. \ref{Sec:Imminence}), that had contained both the necessary ingredients for the scenario of \cite{Garriga_Vilenkin}, and SFS. Taking into account that the resulting model contains a bare minimum of initial assumption, it is reasonable to expect that the qualitative estimates made in it would also hold for more realistic models.

\item Most interestingly, when the model was equipped with the current observational data (see Sec \ref{Sec:Imminence}), it had predicted the imminent arrival of SFS, for its time of emergence ended up being comparable with the age of observable universe (about 10 Gyr).
\end{enumerate}

It is important to note that the models of the universe undergoing a relatively abrupt (that is, compared to the trillions of years) halt in accelerated expansion and collapsing thereafter are not unusual in the framework of the supergravity theory. For example, it has been demonstrated in  \cite{Linde-2}, \cite{Linde-3} that in such cosmological models the dark energy's density eventually evolves to negative values, after which the universe totally collapses in time comparable with the observable age of the universe. An even more general class of such models was considered in \cite{Linde-4}, where it has been shown that any {\em eternally} expanding model of a dark energy universe must have a number of counterparts, where the observable expansion reverts to a contraction in characteristic time frame of  $10^{10}-10^{11}$ years. And while those ``alternative'' models might appear too complex and not very natural, we have to take them into account nevertheless, if for nothing else, but for a simple fact that all of them by design satisfy the current observations (see also \cite{Temchik}).

Getting back to the models with SFS, we would also like to mention the article \cite{Kam-1}, where the authors, while studying the models with a Big Brake singularity (introduced in \cite{Kam-2}), endeavoured to compare the pre-singular evolution of such universes with the supernovae data, and have managed to select some sets of initial parameters compatible with these data. The authors then have numerically analysed the duration of time that has to pass between the moment of (contemporary) observation and the Big Brake singularity, and have shown it to be not very large.

In conclusion, we cannot resist mentioning one last remarkable circumstance. In 2007 the authors of paper \cite{Dab} have made an attempt to estimate a possible time of arrival of SFS using the observational data on Ia type supernovae. They have arrived to an astonishing conclusion that ``a sudden  singularity  may  happen  in  the  very  near future (e.g.  within ten million years)''! It is important to point out that the cosmological model used in \cite{Dab} was very different from the one studied in this paper. And yet the conclusions of  \cite{Dab} and ours are quite alike. Most fascinating, isn't it?

\section*{Acknowledgement} \addcontentsline{toc}{section}{Acknowledgement}
The work was supported from the Russian Academic Excellence Project at the Immanuel Kant Baltic Federal University, and by the project 1.4539.2017/8.9 (MES, Russia). The authors would also like to express their gratitude to anonymous referee for substantial critique of the original version of this article, which led to a significant improvement of the manuscript.


\end{document}